\documentclass[aps,twocolumn,showpacs,superscriptaddress]{revtex4-1}
\usepackage{graphicx,color}
\usepackage{amsmath}
\usepackage{amssymb}
\usepackage{amsfonts}
\usepackage{bm}

\newcommand{\ot}{{\,\otimes\,}}
\newcommand{{\Cd}}{{\mathbb{C}^d}}

\def\A{\mathcal{A}}
\def\B{\mathcal{B}}

\def\i{\mathrm{id}}

\def\i{\mathrm{id}}
\def\g{\mathrm{guess}}
\def\Q{\mathcal{Q}}

\def\oper{{\mathchoice{\rm 1\mskip-4mu l}{\rm 1\mskip-4mu l}%
{\rm 1\mskip-4.5mu l}{\rm 1\mskip-5mu l}}}
\def\<{\langle}
\def\>{\rangle}
\newtheorem{Theorem}{Theorem}
\newtheorem{Lemma}{Lemma}

\newtheorem{Proposition}{Proposition}
\def\tr{\mbox{tr}}
\def\bea{\begin{eqnarray}}
\def\eea{\end{eqnarray}}
\renewcommand{\H}{\mathcal{H}}

\begin{document}

%\title{ Distinguishability Can Characterise (Non-)Markovianity of Quantum Evolution \\
%\RED{Non-Markovian Evolution vs. Distinguishability  of Quantum Channels}}

\title{  Operational Characterization of Divisibility of Dynamical Maps}%: Strict Hierarchy  }
% on Markovianity}

\author{ Joonwoo Bae  }
\affiliation{ Department of Applied Mathematics, Hanyang University (ERICA), 55 Hanyangdaehak-ro, Ansan, Gyeonggi-do, 426-791, Korea, }
\affiliation{ Freiburg Institute for Advanced Studies (FRIAS), Albert-Ludwigs University of Freiburg, Albertstrasse 19, 79104 Freiburg, Germany, }

\author{ Dariusz Chru\'sci\'nski }
\affiliation{Institute of Physics, Faculty of Physics, Astronomy, and Informatics, Nicolaus Copernicus University, Grudziadzka 5, 87-100 Torun, Poland.  }

\date{\today}

%%%%%%%%%%%% Abstract %%%%%%%%%%%%%%%%%%%%%%%%%%%
\begin{abstract}
In this work, we show the operational characterization to divisibility of dynamical maps in terms of distinguishability of quantum channels. It is proven that distinguishability of any pair of quantum channels does not increase under divisible maps, in which the full hierarchy of divisibility is isomorphic to the structure of entanglement between system and environment. This shows that i) channel distinguishability is the operational quantity signifying (detecting) divisibility (indivisibility) of dynamical maps and ii) the decision problem for divisibility of maps is as hard as the separability problem in entanglement theory. We also provide the information-theoretic characterisation to divisibility of maps with conditional min-entropy. 

\end{abstract}


%\pacs{03.65.Ud, 03.67.Hk, 03.65.Wj}

%\begin{document}

%\pacs{03.65.Yz, 03.65.Ta, 42.50.Lc}

\maketitle

Quantum information theory leads to information applications from the principles of quantum theory \cite{QIT,Wilde}. Quantum states are identified as non-replicable resources having cryptographic applications and equivalently, entanglement, arguably a general resource for quantum information applications, contains monogamous correlations. Quantum channels, one of the most fundamental ingredient of quantum information theory, describe legitimate dynamics of quantum resources, i.e., a dynamical map $\Lambda_t$ from initial state $\rho$ to resulting one at time $t$, $\rho_t = \Lambda_t [\rho]$. They can be realized as, in general,
\begin{equation}\label{}
  \Lambda_t[\rho] = {\rm Tr}_E ( U^{SE}_t \rho \ot \rho_E U^{SE \dagger}_t) \label{eq:1} ,
\end{equation}
with unitary transformation $U^{SE}_t$ acting on $\mathcal{H}^{(S)}  \ot \mathcal{H}^{(E)}$ and a fixed state of the environment $\rho_E$ living in  $\mathcal{H}^{(E)}$. Interacting with environment under evolution not in the form $U_{t}^{S} \otimes U_{t}^{E}$, the system must be treated as an open quantum system \cite{Breuer}.

Open quantum systems and their dynamical features have been intensively studied. They are not only of a general interest as the description, see Eq. (\ref{eq:1}), has no classical counterpart in the fundamental point of view, but also crucial for the analysis of the system-environment interaction which is responsible for system dissipation, decay, and decoherence \cite{Breuer,Weiss,RIVAS,Alicki}. It should be stressed that the robustness of quantum coherence and entanglement against the destructive  effects of the environment is essential for variety of applications of quantum physics in modern quantum technologies due to the fact that both quantum entanglement and quantum coherence are basic resources for quantum information processing \cite{QIT}. 

Recently, much effort has been devoted in particular to the description, analysis, and classification of non-Markovian evolution (see e.g. \cite{ref:review,ref:review2} for the recent review papers  and the collection of papers in \cite{JPB}). The most popular approaches exploit distinguishability of states  \cite{BLP}, CP-divisibility of dynamical maps \cite{Wolf,RHP,Wolf-Isert, BN}, quantum Fisher information flow \cite{Fisher}, fidelity \cite{fidelity}, mutual information \cite{Luo1,Luo2}, channel capacity \cite{Bogna},  geometry of the set of accessible states \cite{Pater}, and  quantum interferometric power \cite{Adesso}.

Among these approaches distinguishability of states leads to an operational characterisation of the so-called P-divisibility, where a dynamical map can be decomposed such that  $\Lambda_t = \Lambda_{t,s} \circ \Lambda_s$ as a concatenation of a positive map $\Lambda_{t,s}$ and a quantum channel $\Lambda_s$ for all $t\geq s \geq 0$. In fact, P-divisibility is directly related to Markovianity of classical evolution, which is to be discussed later. Remarkably, the approach identifies the notion of information that signifies Markovianity, by introducing {\it information flow} that is the time-rate of state distinguishability \cite{BLP}. Let us recall that $\Lambda_t$ is called CP-divisible if the family of maps $\Lambda_{t,s}$ is completely positive \cite{Wolf,RHP,Wolf-Isert,Hou}. CP-divisibility is a resource in quantum communication, e.g. quantum subdivision code \cite{ref:wolf}. Interestingly, one may introduce the notion of $k$-divisible map $\Lambda_t$ for which $\Lambda_{t,s}$  is $k$-positive  \cite{ref:CM14}. This refinement covers all the spectrum from P- to CP-divisibility $k=1,2,\ldots,d={\rm dim \H}^{(S)}$.

In this paper, we show the operational characterisation to $k$-divisibility of dynamical maps using the unifying idea of quantum channel discrimination. This merges different approaches of characterizing Markovianity. We show that for any pair of quantum channels, their distinguishability does not increase under divisible maps, and conversely, indivisible maps are detected by finding an infinitesimal increase of distinguishability for some pair of channels. Our results establish the isomorphic relations among divisibility of dynamical maps, entanglement between system and environment, and distinguishability of channels, from which the interaction between system and environment is found in the operational way with the view of entanglement theory. We also present an entropic chararcterization to divisibility of maps in terms of min-entropy.

We begin with Markovian process in classical information theory, where Markovianity has been a well-established concept \cite{Feller,Kampen}.  A classical stochastic process $x_i(t)$ is called Markovian if the conditional probability defining the process
satisfies \cite{Kampen}
\bea
p(x_i,t_i | x_{i-1}, t_{i-1} ; \cdots;  x_1,t_1 ) = p(x_i,t_i | x_{i-1}, t_{i-1} ) , \nonumber
\eea
for $t_i\geq t_{i-1} \geq \cdots \geq t_1$. This property means essentially that the processes has no memory about the past events. It implies that $p(x,t|y,s)$ for time $t>s$ satisfies the Chapman-Kolmogorov equation
\begin{equation}\label{CK}
  p(x,t|y,s) = \sum_z   p(x,t|z,u) \, p(z,u|y,s)\ ,
\end{equation}
for any $u$ satisfying $t > u > s$.  This concept, however, cannot be simply transferred into the quantum domain \cite{ref:BLP11}.

Let us now consider linear evolution of states in general, that is, probability vectors and density operators in the classical and quantum case, respectively. In the classical case, a probability vector is denoted by $\mathbf{p}_t$ and then Markovian evolution is represented by a P-divisible dynamical map, i.e. a family of stochastic matrices $T(t)$ satisfying
\begin{equation}\label{}
T(t) = T(t,s) T(s) ,
\end{equation}
where $T(t,s)$ defines a stochastic matrix for any $t\geq s \geq 0$. Note that $T(t,s)$ satisfy the Chapman-Kolmogorov equation $T(t,s) =T(t,u)T(u,s)$ which expresses the Markovianity of the evolution, see the discussion \cite{ref:BLP11}. This property is fully characterized in terms of time-local Kolmogorov generator \cite{Kampen}
\begin{equation}\label{LG}
  \frac{d}{dt} {T}(t) = K(t) T(t) \ ,
\end{equation}
that is, $K_{ij}(t) \geq 0$ for $i \neq j$, and $\sum_i K_{ij}(t) = 0$.

The quantum analog corresponds to a dynamical map $\Lambda_t$ for $t \geq 0$. Similarly to the classical case, one may identify the quantum evolution represented by $\Lambda_t$ as Markovian if and only if the corresponding dynamical map is a concatenation of two quantum channels as follows,
\begin{equation}\label{eq:cpd}
  \Lambda_t = \Lambda_{t,s} \circ \Lambda_s
\end{equation}
with some legitimate quantum channel $\Lambda_{t,s}$ defined for $t \geq s \geq 0$. This property is called CP-divisibility and implies that $\Lambda_{t,s} =  \Lambda_{t,u} \circ \Lambda_{u,s}$, which may be regarded as quantum analog of the Kolmogorov-Chapman equation in Eq. (\ref{CK}).  Analogously to the classical case in Eq. (\ref{LG}), CP-divisibility is fully characterized on the level of time-local generators
\begin{equation}\label{}
   \frac{d}{dt} {\Lambda}_t = L_t \Lambda_t ,
\end{equation}
where $L_t$ has the following well known structure \cite{GKS, L}
\bea
L_t[\rho]  & = & - i [H(t),\rho]  \nonumber\\
&& + \sum_{\alpha=1}^n \gamma_{\alpha} (t) \left[ A_{\alpha} (t) \rho A_{\alpha}^{\dagger} (t)  -  \frac{1}{2} \{ A_{\alpha}^{\dagger} (t) A_{\alpha} (t), \rho \} \right] \nonumber %~~\label{eq:master}
\eea
with time-dependent relaxation rates $\gamma_{\alpha}(t) \geq 0$ and noise operators $A_{\alpha}(t)$.

The notion of divisibility has been refined to $k$-divisibility very recently \cite{ref:CM14}. Namely, a dynamical map $\Lambda_t$ is called $k$-divisible if and only if the propagator $\Lambda_{t,s}$ in Eq. (\ref{eq:cpd}) is $k$-positive, i.e., it remains positive when extended to $k$-dimensional ancilla: $\oper_k \ot \Lambda_t$ is positive. The case $k=1$ corresponds to P-divisibility, and the other $k= d$ $(=\mathrm{dim} \H^{(S)} )$ to CP-divisibility. Then, the notion of $k$-divisibility scans from P- to CP-divisibility over dynamical maps, classifying the depth of interaction between system and environment.

{\em Channel discrimination}.  We next move to describe channel discrimination, and show the fine structure in the quantum channel discrimination in its relation to entanglement theory. Let us begin with a stochastic process as follow. For an event space $E$, suppose that there are two distributions $\mathbf{p}_1 (x)$ and $\mathbf{p}_2 (x)$ for $x\in E$, and each of them appears with {\it a priori} probabilities $1-p$ and $p$, respectively. We recall the variational distance of the distributions, denoted by $D^p[\mathbf{p}_1,\mathbf{p}_2] = \sum_{x\in E} | (1-p) \mathbf{p}_1 (x) - p  \mathbf{p}_2 (x) | $, which we call $p$-distinguishability for probabilistic systems, quantifies the success probability of making correct guess about the distributions, $p_{\g} =  ( 1 + D^p [\mathbf{p}_1,\mathbf{p}_2] )/2$. Instead of distributions, one can consider a channel that is applied probabilistically, either of two stochastic matrices $S_1$ or $S_2$ with probabilities $1-p$ or $p$. Once a probability measure $\mathbf{p}$ is taken, it evolves to either $\mathbf{p}_1 = S_1\mathbf{p}$ or $\mathbf{p}_2 = S_2\mathbf{p}$. Optimal discrimination between the channels leads to finding an optimal probability measure $\mathbf{p}$ that maximizes the variational distance of resulting distributions as follows,
\bea
D_c^p[S_1,S_2] = \max_{\mathbf{p} } \sum_{x \in E} | (1-p) S_1 \mathbf{p} (x) - p  S_2\mathbf{p} (x) | \nonumber
\eea
which we call $p$-distinguishability for classical channels. Note that the guessing probability about channels is given by $p_{\g} = (1+D_{c}^p[S_1 , S_2])/2$. It is worth to mention that the variational distance of distributions has the operational meaning as distinguishability, and also that for classical cases, channel discrimination is essentially equivalent to state discrimination. A classical process is well-defined as stochastic mappings over distributions solely on systems. 

Analogously to the variational distance of probability distributions, the trace norm defined for a hermitian operator $X$ as $\| X \|_{\rm tr} = \tr \sqrt{X X^\dagger}$ quantifies distinguishability for quantum states \cite{ref:helstrom}, that is, $p$-distinguishability of two density operators $\rho_1$ and $\rho_2$ reads $D^p[\rho_1,\rho_2]=||(1-p)\rho_1 - p \rho_2||_{\rm tr}$. One may consider distinguishability of quantum channels only by discriminating between output states resulting from the channels, as it is done in the above. However, contrary to the classical case, a system and ancillas can be in entangled states, that indeed leads to the improvement in quantum channel discrimination \cite{ref:piani}. Here, we make even finer approach that takes into account the degrees of freedom of ancillas. In what follows, we show that there exists a strict hierarchy of channel distinguishability according to entanglement between system and ancillas: namely, the more entangled system and acillas are the more useful they are for channel discrimination.

\begin{figure}
\begin{center}  % Requires \usepackage{graphicx}
\includegraphics[width= 8.5cm]{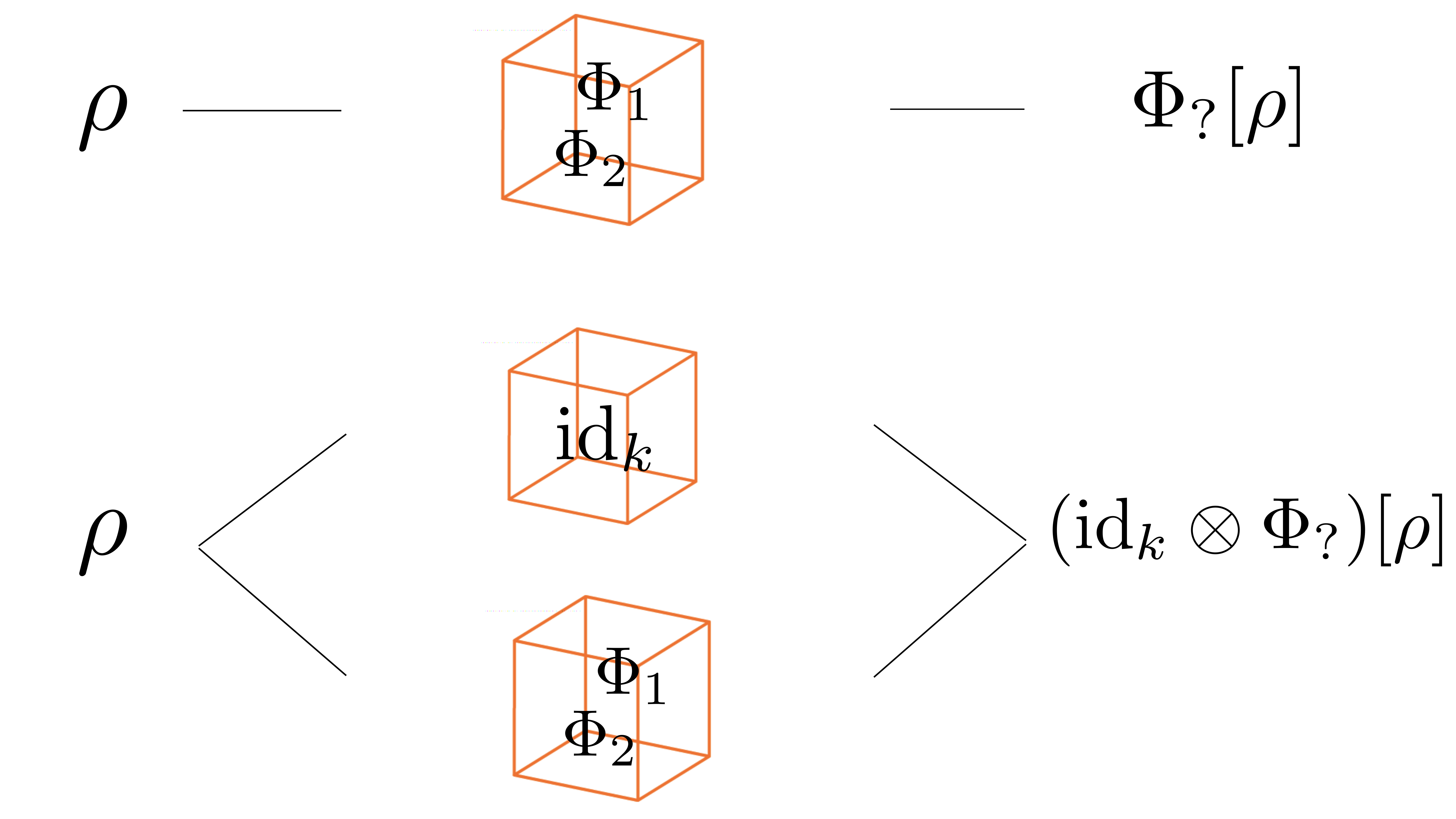}
\caption{ A quantum channel $\Phi_1$ or $\Phi_2$ with $k$-dimensional ancillas is applied to an input state $\rho$ with {\it a priori} probabilities, and can be discriminated from resulting states $(\mathrm{id}_k \otimes \Phi_{?})[\rho]$. (A) For $k=1$, it is equivalent to state discrimination between $\Phi_1 \rho$ and $\Phi_2 \rho$, and (B) for $k>1$, channel distinguishability in Eq. (\ref{eq:disk}) can be improved with entangled states $\rho$.}
\end{center}
\end{figure}

To begin with, we define the trace norm of a Hermitian map $\Psi$ that acts on systems in $\H^{(S)}$ while $k$-dimensional ancillas in $\mathcal{H}_k^{(A)}$ are assisted, as follows,
\begin{equation}\label{}
||\Psi||^{(k)}_{\rm tr} = \max_{\rho_{AS}} || \oper_k \ot \Psi[\rho_{AS}]||_{\rm tr} \label{eq:nok}
\end{equation}
where the maximization runs over all quantum states $\rho_{AS}$ in $\mathcal{H}_k^{(A)} \ot \mathcal{H}^{(S)}$. The case of $k=1$ means that no ancilla is applied. It also suffices consider $k$ up to the system's dimension, $k\leq d$ $(=\mathrm{dim}\H^{(S)})$, with the equality that corresponds to the norm of complete boundness or known as diamond norm denoted by $\|\cdot\|_{\diamond}$ \cite{ref:kitaev}.

Consider now the channel discrimination problem that channels $\Phi_1$ and $\Phi_2$ are given with {\it a priori} probabilities $1-p$ and $p$, respectively. We write the trace distance of the channels as,
\bea
D_{k}^{p} [\Phi_1 , \Phi_2] &:=&  \| ((1-p)  \Phi_1 - p \Phi_2  \|_{\rm tr}^{(k)} \nonumber \\
& =& \max_{\rho_{AS}}  \|  \oper_k \otimes ( (1-p)  \Phi_1 - p  \Phi_2) [\rho_{AS}] \|_{\rm tr} ~~~\label{eq:disk}
\eea
with optimization over input states, and we call the distance as $p$-distinguishability of quantum channels. This rephrases $p$-distinguishability of two states $\oper_k \otimes \Phi_1 [\rho_{AS}]$ and $\oper_k \otimes \Phi_2 [\rho_{AS}]$ given with probabilities $1-p$ and $p$ respectively. Then, the guessing probability for channels is given as, $p_{\g} =  ( 1 + D_k^p [\Phi_1,\Phi_2] ) /2$. We note that the distance measure has the operational meaning as distinguishability of channels.

One can notice a natural chain of inequalities for the norm of a Hermitian map $\Psi$ in Eq. (\ref{eq:nok}):
\begin{equation}\label{}
  \| \Psi \|_{\rm tr} =  \| \Psi \|_{\rm tr}^{(1)} \leq  \ldots \leq \| \Psi \|^{(d-1)}_{\rm tr}  \leq   \| \Psi \|_{\rm tr}^{(d)} = \|\Psi \|_\diamond . \nonumber
\end{equation}
Consequently, for distinguishability of channels it follows that any pair of channels $\Phi_1$ and $\Phi_2$, and $p\in[0,1]$, we have
\begin{equation*}\label{}
 D^{p} [\Phi_1 , \Phi_2] =
 D_{1}^{p} [\Phi_1 , \Phi_2] \leq D_{2}^{p} [\Phi_1 , \Phi_2] \leq \ldots \leq D_{d}^{p} [\Phi_1 , \Phi_2], \nonumber
\end{equation*}
where $D^{p} [\Phi_1 , \Phi_2]$ is denoted for the case $k=1$ for convenience.
%\RED{
It has been proved \cite{ref:piani} that a state $\rho$ living in $\mathcal{H} \ot \mathcal{H}$ is entangled if and only if there exist channels $\Phi_1$ and $\Phi_2$ such that
\begin{equation}\label{}
  \| (1-p)\oper \ot \Phi_1[\rho] - p \oper \ot \Phi_2[\rho] \|_{\rm tr} >  D_{1}^{p} [\Phi_1 , \Phi_2] \ ,
\end{equation}
for $p= 1/2$. Actually, the result can be generalized for arbitrary $p$. It shows the essential role entanglement plays in a channel discrimination problem. %}

Let us take the entanglement measure, Schmidt number, to quantify entanglement between system and ancillas. Let $\mathrm{SR}(|\psi\rangle)$ denote the Schmidt rank of a pure state $|\psi\rangle$, which is the number of non-vanishing coefficients in the Schmidt decomposition. For a mixed state $\rho$ in general, it can be extended by the convex-roof construction and is called Schmidt number: $\mathrm{SN}(\rho) = \min_{\{p_k , |\psi_k\rangle \} } ( \max_{k} \mathrm{SR}(|\psi_k\rangle) )$ with minimization over all decompositions. Consequently, Schmidt number is an entanglement measure. We write by $\Q_k$ the set of quantum states having their Schmidt number no greater than $k$. Then, there exists a natural chain of the subsets 
\bea
\Q_1 \subset \Q_2 \subset \ldots \subset \Q_d, \label{eq:shier}
\eea
where one can observe that $\Q_1$ corresponds to the set of separable states and $Q_d$ to all quantum states in $\mathcal{H}_d \ot \mathcal{H}_d$. It is well-known that $\rho \in \Q_k$ if and only if $\oper_k\otimes\Lambda [\rho]\geq 0$ for all $k$-positive maps $\Lambda$. All these are to be used to find the hierarchical structure in channel distinguishability according to dimension $k$ of ancillas in such a way that it is isomorphic to the entanglement monotone, Schmidt number, over quantum states.

\begin{figure}
\begin{center}   
\includegraphics[width= 9cm]{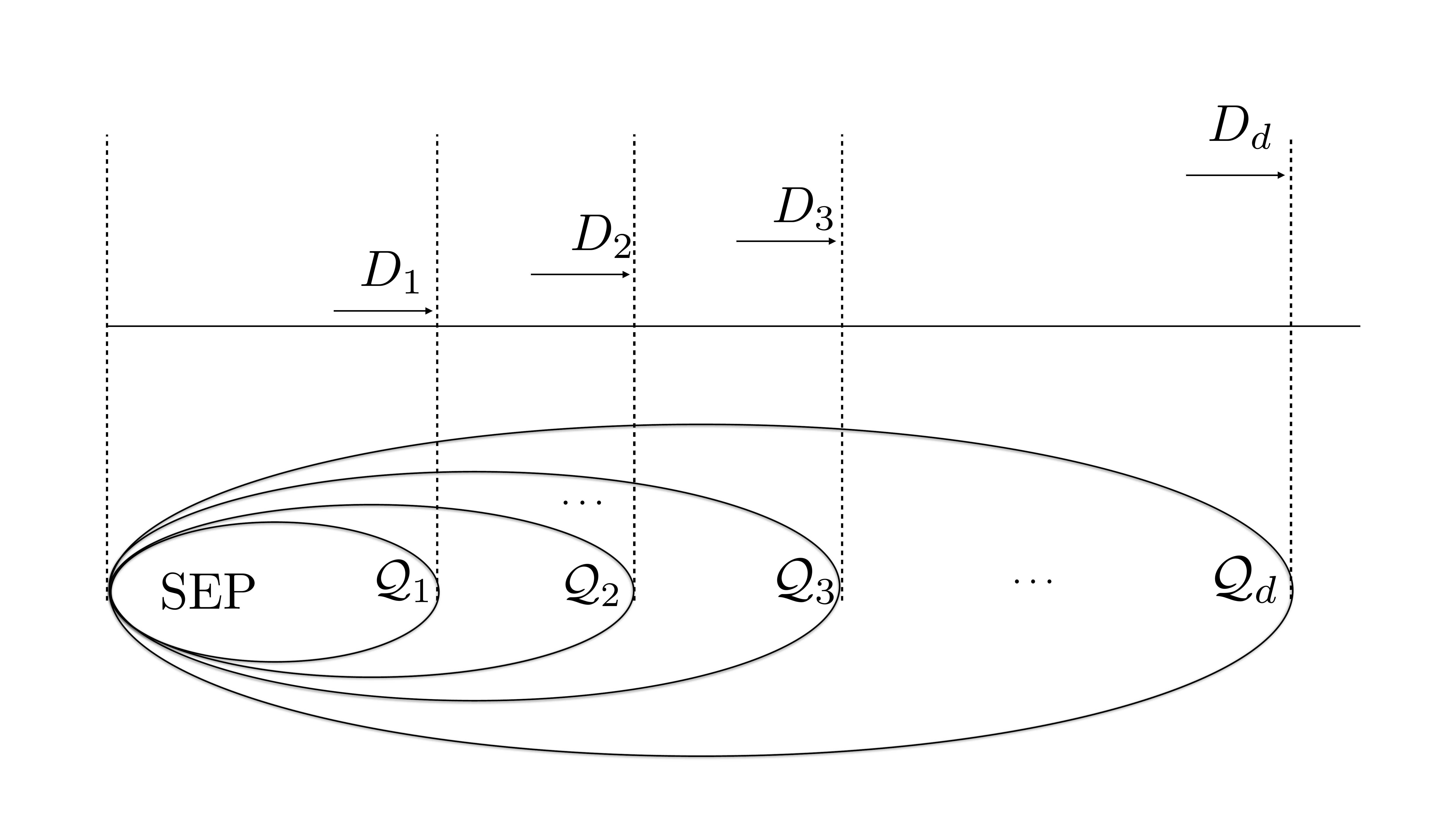}
\caption{ A structural hierarchy exists among quantum states in terms of an entanglement measure, Schmidt number, see Eq. (\ref{eq:shier}). Distinguishability $D_k$ in Eq. (\ref{eq:disk}) is structured according to entanglement between system and ancillas. }
\end{center}
\end{figure}
 
{\em Characterizing divisibility of maps}. With the hierarchical structure in channel distinguishability, we now present the operational characterization to divisibility of dynamical maps in terms of $p$-distinguishability, and readily find the strict hierarchy among divisible maps. Let us first revisit the classical case, stochastic evolution of the probability vector $\mathbf{p}_t$. It should be clear that in the classical setting discrimination of channels is essentially equivalent to discrimination of states. Denoted by $\mathbf{p}_k = S_k \mathbf{p}$ for channels $S_k$ with $k=1,2$ and some probability vector $\mathbf{p}$, classical evolution $T_t$ is P-divisible, i.e. Markovian, if and only if
\begin{equation}\label{C}
\frac{d}{dt} D_c^{p}[T_t S_1,T_t S_2] \leq 0 ,
\end{equation}
for any two classical  channels $S_1,S_2$ and $0 < p < 1$.

The result can be immediately generalized to the quantum domain, that includes the so-called Breuer-Laine-Piilo (BLP) definition of Markovianity \cite{BLP}. Assuming that a dynamical map $\Lambda_t$ is invertible, we have the following.
\begin{Proposition} Quantum evolution $\Lambda_t$ is P-divisible if and only if,
\begin{equation}\label{Q-1}
 \frac{d}{dt} D^p[\Lambda_t \circ \Phi_1,\Lambda_t \circ \Phi_2] \leq 0 ,
\end{equation}
for any pair of quantum channels $\Phi_1$ and $\Phi_2$.
\end{Proposition}

In the above, $p$-distinguishability of channels is equivalent to $p$-distinguishability of states by relating channels and states as $\rho_k = \Phi_k[\rho]$ one can replace Eq. (\ref{Q-1}). It should be clear that Eq. (\ref{Q-1}) with $p=1 /2$ is equivalent to the BLP definition of Markovianity. The equivalence between state and channel discrimination, however, no longer holds true if one is allowed to exploit ancillas, since entangled states over system and ancillas can improve distinguishability of quantum channels \cite{ref:kitaev,ref:mauro,ref:piani}. This in fact makes it highly non-trivial to derive the definition of Markovian quantum evolution. All these can be encapsulated into the refined notion of $k$-divisibility: see Eq. (\ref{eq:cpd}), a map $\Lambda_t$ is $k$-divisible if it can be divided as $\Lambda_t = \Lambda_{t,s}\circ \Lambda_{s}$ with $k$-positive map $\Lambda_{t,s}$ for $t \geq s\geq 0$. 

We are now ready to state the main result, the operational characterization to $k$-divisible maps with distinguishability of channels.
\begin{Theorem}
Quantum evolution $\Lambda_t$ is $k$-divisible if and only if, for any $0<p<1$,
\begin{equation}\label{Q}
\frac{d}{dt} D^p_k[\Lambda_t \circ \Phi_1,\Lambda_t \circ \Phi_2] \leq 0 ,
\end{equation}
for any pair of quantum channels $\Phi_1$ and $\Phi_2$.
\end{Theorem}
This finds the operational characterization for divisible maps in general: a map that is not $k$-divisible can be detected by observing increase of $p$-distinguishability of channels with $k$-dimensional ancillas, for some $p\in(0,1)$. We remark that the operational characterization leads to devising experimental schemes of detecting indivisible maps by finding time-rate of distinguishability, even without knowing a map itself precisely. This resembles to entanglement witnesses that can detect entanglement of unknown states.
 
{\em Information-theoretic characterisation.} In the following, we present the information-theoretic characterization of divisible maps. This exploits the link between distinguishability and the quantum conditional min-entropy \cite{ref:KRS}. To relate channel distinguishability with min-entropy, let us suppose a scenario that two parties, Alice and Bob, play a guessing game as follows. Bob first prepares a quantum state $\rho$ in $\H_{k}^{(A)} \otimes \H^{(S)}$ of $k$-dimensional ancilla and system, and sends it to Alice. She then applies one of quantum channels $\mathrm{id}_k\otimes \Phi_1$ and $\mathrm{id}_k\otimes \Phi_2$ with probabilities $1-q$ and $q$, respectively. She registers her application as $| i \rangle \langle i|_A$ for $i=1,2$. The resulting state returns to Bob, where system evolves under a dynamical map $\Lambda_t$ over state space $S(\H^{(S)})$. The shared state contains classical-quantum (cq) correlations at time $t$ as follows,
\bea
\rho_{A B_k}(t) = \sum_{i=1,2} q_i   |i\rangle \langle i|_A \otimes ( \mathrm{id}_k \otimes \Lambda_t  \circ \Phi_i ) [\rho]_B.  \label{eq:cqchannel}
\eea
where $q_1 =1-q$ and $q_2=q$. Given the cq correlations, Bob's maximal information about Alice's preparation, that is, application of channels, followed by his system evolution $\Lambda_t$, is quantified by conditional min-entropy and in fact achieved by optimal discrimination between resulting states. That is, it holds that $ H_{\min} (A | B)_{\rho_{AB_k} (t)} = -\log p_{\g}(A|B)_{\rho_{AB_k}(t)} $ for cq correlations $\rho_{AB_k}(t)$ in Eq. (\ref{eq:cqchannel}) and the guessing probability is equivalent to $q$-distinguishability of quantum channels for $\Phi_1$ and $\Phi_2$. Thus, we have
\bea
H_{\min} (A | B )_{\rho_{A B_k }(t)} = - \log \frac{1}{2}  (1+ D_{k}^{q} [ \Lambda_t \circ \Phi_1, \Lambda_t  \circ \Phi_2 ] ) ~~~~~~\label{eq:cq}
\eea
From the relation in the above and the characterization in Eq. (\ref{Q}), we have the information-theoretic characterization to $k$-divisibility of quantum channels as follows.
\begin{Proposition} A dynamical map $\Lambda_t$ is $k$-divisible if and only if for classical-quantum correlation $\rho_{AB_k} (t)$ in Eq. (\ref{eq:cqchannel}) the conditional min-entropy does not decrease, i.e.,
\bea
\frac{d}{dt} H_{\min}(A  | B )_{\rho_{AB_k} (t)} \geq 0 \nonumber
\eea
for any pair of quantum channels $\Phi_1$ and $\Phi_2$.
\end{Proposition}
That is to say, indivisibility of dynamical map including non-Markovianity is indicated by decrease of min-entropy for some pair of channels. Similarly to the second law of thermodynamics, the conditional min-entropy of any pair of channels never decreases under divisible maps.

In conclusion, we have provided the operational characterization to divisibility of quantum channels in terms of quantum channel discrimination. The characterization takes into account the effective size of ancillas, or equivalently effective degrees of freedom in environment, entangled with a given system. For any pair of channels, their distinguishability with $k$-dimensional ancillas does not increase under $k$-divisible dynamical maps. This is the necessary and sufficient condition for divisibility of maps. All these are connected to the hierarchy of entanglement between system and environment. Distinguishability of channels is an operational quantity that can be measured in practice, by which divisibility of maps and entanglement of system and environment can be detected. 

We have extended the analysis to the information-theoretic characterization, and have shown that similarly to the second law of thermodynamics, quantum conditional min-entropy of quantum channels never decreases under divisible maps. Our results establishes the correspondence between channel distinguishability and divisibility via entanglement theory, and derive operational and information-theoretic characterizations to divisibility of dynamical maps. This would lead to methods of detecting indivisible dynamical maps in a similar way of non-Markovianity witnesses \cite{ref:witness}, and be applied to deriving quantifications such as indivisibility measures as well as non-Markovianity, see e.g. \cite{BLP}. Our work may envisage useful and fundamental understanding on defining Markov processes with the tool of distinguishability. Moreover, the isomorphic connection between hierarchies of divisible maps and entanglement implies immediately that divisibility is a useful classification of maps, such as a resource theory of maps.

In recent years, it is found that divisibility is closely related to other fundamental aspects of quantum theory. It turns out that CP divisibility is closely related to the so-called temporal quantum steering \cite{ref:chen}, a weaker form of the Leggett-Garg (LG) inequality devised originally for the macroscopic realism \cite{ref:lg}. However, the LG inequality itself cannot provide the tight characterization for the purpose \cite{ref:kofler}, contrast to the case of local realism \cite{ref:bell}. On the other hand, {\it all} physical theories, both macroscopic realism and local realism, fulfill the no-signaling principle that is tightly connected to distinguishability of quantum states \cite{ref:bae}, which corresponds to P divisibility. It would be interesting to find how $k$-divisibility of dynamical maps structures temporal quantum correlations, or is related to fundamental properties of quantum dynamics in general. Our findings may provide the theoretical framework that opens up a new avenue to investigate fundamental properties of quantum dynamics and their information applications. \\

\section*{Acknowledgement}
This work is supported by the People Programme (Marie Curie Actions) of the European Union Seventh Framework Programme (FP7/2007-2013) under REA grant agreement N. 609305, Institute for Information \& communications Technology Promotion(IITP) grant funded by the Korea government(MSIP) (No.R0190-16-2028, PSQKD) and the National Science Center project 2015/17/B/ST2/02026. The work was done partially while the authors were visiting the Institute for Mathematical Sciences, National University of Singapore in 2013. The visit was supported by the Institute.

\newpage

\newpage

\section*{Appendix}

\subsection*{Proof of the main theorem}

We here provide the proof for the main theorem. To this end, we need the following lemma. \\

\begin{Lemma} If $\Phi: \B(\H) \rightarrow \B(\H)$ is trace-preserving then $\Phi$ is positive if and only if $\|  \Phi(X) \|_1 \leq \|  X \|_1 $ for all $X\in \B(\H)$.
\end{Lemma}

%$Theorem.$ $\Lambda_t$ is $k$-divisible {\bf  if and only if}
%\bea
%\frac{d}{dt} D_{k}^{(q)} [\Lambda_t \circ \Phi_1 , \Lambda_{t} \circ  \Phi_2  ] \leq 0,\label{eq:theorem}
%\eea
%for all quantum channels $\Phi_1$ and $\Phi_2$ and $q\in [0,1]$.
% \\
Proof of Theorem. ($\Rightarrow$) Suppose that a dynamical map $\Lambda_t$ is $k$-divisible, i.e., there exist a class of $k$-positive propagators $\Lambda_{t,s}$ such that the map can be decomposed as
\bea
\Lambda_t = \Lambda_{t,s} \circ \Lambda_s,~\forall~  t\geq s \geq 0. \label{eq:con}
\eea
For any $q\in [0,1]$ and some channels $\Phi_1$ and $\Phi_2$, since $ \Lambda_t$ is $k$-divisible we have that
\bea
&& D_{k}^{[q_1,q_2]} [\Lambda_{t+\epsilon} \circ \Phi_1 , \Lambda_{t+\epsilon} \circ  \Phi_2  ]  \label{eq:exp1}\\
&=& \max_{\rho\in S_k}  \|  \i \otimes \Lambda_{t+\epsilon}\circ (q_1 \Phi_1 - q_2 \Phi_2) )[\rho]   \|  \label{eq:exp2} \\
&= & \max_{\rho\in S_k}  \|  \i  \otimes \Lambda_{t+\epsilon ,t} \circ \Lambda_t \circ (q_1 \Phi_1 - q_2 \Phi_2) )[\rho]   \|  \label{eq:exp3} \\
&\leq & \max_{\rho\in S_k}  \|  \i \otimes \Lambda_t \circ (q_1 \Phi_1 - q_2  \Phi_2) )[\rho]  \|  \label{eq:exp4}  \\
& = & D_{k}^{ [q_1,q_2]} [\Lambda_{t } \circ \Phi_1 , \Lambda_{t } \circ  \Phi_2  ].  \label{eq:exp5}
\eea
The inequality in Eqs. (\ref{eq:exp3}) and (\ref{eq:exp4}) holds since a $k$-positive map is positive over all states having Schmidt ranks no greater than $k$. It thus follows that
\bea
&& \frac{d}{dt} D_{k}^{[q_1,q_2]} [\Lambda_t \circ \Phi_1 , \Lambda_{t} \circ  \Phi_2  ]  \nonumber \\
&=& \lim_{\epsilon\rightarrow +0} \frac{1}{\epsilon}  [ D_{k}^{[q_1,q_2]} [\Lambda_{t+\epsilon} \circ \Phi_1 , \Lambda_{t+\epsilon} \circ  \Phi_2  ]   \nonumber \\
&& ~~~~~~~~ - D_{k}^{ [q_1,q_2]} [\Lambda_t \circ \Phi_1 , \Lambda_{t} \circ  \Phi_2  ]] \nonumber\\
&\leq & \lim_{\epsilon\rightarrow +0} \frac{1}{\epsilon}  [ D_{k}^{ [q_1,q_2]} [\Lambda_{t  } \circ \Phi_1 , \Lambda_{t } \circ  \Phi_2  ]   \nonumber\\
&& ~~~~~~~~ - D_{k}^{[q_1,q_2]} [\Lambda_t \circ \Phi_1 , \Lambda_{t} \circ  \Phi_2  ]]  = 0.  \nonumber
\eea

($\Leftarrow$) Conversely, for all quantum channels and $q\in[0,1]$, it holds that $\forall \epsilon\geq 0$,
\bea
D_{k}^{ [q_1, q_2]} [\Lambda_{t+\epsilon} \circ \Phi_1 , \Lambda_{t+\epsilon} \circ  \Phi_2  ] \leq  D_{k}^{[q_1,q_2]} [\Lambda_{t } \circ \Phi_1 , \Lambda_{t } \circ  \Phi_2  ], \nonumber
\eea
which also means the inequality in Eqs. (\ref{eq:exp3}) and (\ref{eq:exp4}). For channels $\Lambda_t$, $\Phi_1$, $\Phi_2$, and parameters $q_1,~q_2\in[0,1]$, let $\rho_0 \in \B (\H_k \otimes \H)$ denote the state that achieves the maximisation in Eq. (\ref{eq:exp4}), i.e.
\bea
&& \max_{\rho\in S_k}  \|  \i \otimes \Lambda_t \circ (q_1 \Phi_1 - q_2\Phi_2) )[  \rho]  \| \nonumber \\
& = &  \max_{\rho \in \B(\H_k \otimes \H) }  \|  \i_k \otimes \Lambda_t \circ (q_1 \Phi_1 - q_2 \Phi_2) )[ \rho] \|  \nonumber\\
& = &   \|  \i_k \otimes \Lambda_t \circ (q_1 \Phi_1 - q_2 \Phi_2) )[  \rho_0]  \|
\eea
Denoted by $Y = \i_k \otimes \Lambda_t \circ (q_1 \Phi_1 -q_2 \Phi_2 )[\rho_0]$, the inequality in Eqs. (\ref{eq:exp3}) and (\ref{eq:exp4}) implies that for all $Y$ given by channels,
\bea
\|  \i_k \otimes \Lambda_{t + \epsilon, t} [Y] \| \leq \| Y \|. \nonumber
\eea
Thus, it is shown that $\Lambda_{t+\epsilon , t}$ is $k$-positive. \hfill $\Box$

%\section*{Appendix II.  Equivalence of Eqs. (\ref{eq:chndis}) and (\ref{eq:chndis2})}

%Note that a $k$-positive map $V_k$

\vspace{.5cm}

\subsection*{Mixing does not increase distinguishability }

We here show that quantum states attaining optimal channel discrimination are pure states. For channels $\{q_i,\Phi_i \}_{i=1}^2$, we write by $\widetilde{\Phi} = q_1 \Phi_1 - q_2 \Phi_2$. Suppose that a state $\rho$ has has a pure-state decomposition as $\rho = \sum_{i} p_i |\psi_i \rangle\langle \psi_i | $. Let $\A_i^{\pm}$ denote positive and negative projections in the decomposition of $( \i_k\otimes \widetilde{\Phi} ) [| \psi_i \rangle \langle \psi_i |]$, so that
\bea
[( \i_k\otimes \widetilde{\Phi} ) [| \psi_i \rangle \langle \psi_i |] ] = \A_i^{+ } - \A_i^{-},~~\mathrm{with}~ \A_i^{\pm} \geq 0. \nonumber
\eea
Then, it holds that
\bea
\| ( \i_k \otimes \widetilde{\Phi} ) [\rho] \|_1  & = &   \|  \sum_{i} p_i (\i_k \otimes \widetilde{\Phi}  ) [|\psi_i\rangle \langle \psi_i | ]  \|_1 \nonumber\\
&= &  \| \sum_{i} p_i (  \A_i^{+ } -   \A_i^{-}  ) \|_1  \nonumber \\
&\leq &  \sum_{i} p_i ( \|  \A_i^{+ } \|_1 + \| \A_i^{-}   \|_1 )  \nonumber \\
&\leq & \max_i  ( \|  \A_i^{+ } \|_1 + \| \A_i^{-}   \|_1 ) \nonumber\\
&= & \max_i  \| ( \i_k \otimes \widetilde{\Phi} ) [ | \psi_i \rangle \langle \psi_i |  ] \|_1.  \nonumber
\eea

\end{document}